\documentclass[twocolumn,showpacs,amssymb,aps]{revtex4}

\usepackage{graphicx}
\usepackage{psfig}
\usepackage{dcolumn}
\usepackage{bm}

\newcommand{\ave}[1]{\left\langle #1 \right\rangle}
\def\lessim{\lower.5ex\hbox{$\; \buildrel < \over \sim \;$}}
\begin{document} \hbadness=10000
\topmargin -0.8cm\oddsidemargin = -0.7cm\evensidemargin = -0.7cm
\preprint{}

\title{Fluctuations in jet momentum as an energy loss probe}
\author{Giorgio Torrieri, Charles Gale, Sangyong Jeon, Vasile Topor Pop}
\affiliation{
Department of Physics, McGill University, Montreal, QC H3A-2T8, Canada}

\date{February, 2006}

\begin{abstract}
We investigate the measurement of the distribution of away-side jet $p_T$ as
a way to probe the energy loss mechanism in heavy ion collisions, and
constrain the properties of the medium created in heavy ion collisions.
We define an observable related to the fluctuation of the energy deposited in the medium, and show that
competing models of parton energy loss give differing values and scaling of this observable.  We also argue that the scaling of this observable with
system size can be related to the medium's partonic density. We then give a qualitative discussion of how the measurement we suggest
can be performed and used to determine the parton energy loss mechanism and possibly the system's partonic density.
\end{abstract}

\maketitle
Jet suppression has long been regarded as a promising signature for deconfinement/Quark Gluon Plasma (QGP) production.  It is very reasonable to suppose that partonic energy loss in a deconfined medium is substantially greater than a medium where
quarks and gluons are locked in color-neutral bags \cite{loss1}.

Perturbative Quantum Chromodynamics (pQCD) can be used to analyze jet energy loss quantitatively, and to explore the features that suppression in a deconfined medium should exhibit \cite{loss2,loss3,bdmps1,loss5}.

The convincing observation of jet suppression in RHIC A-A collisions, and the determination, via d-Au control measurements, that in-medium properties rather than  initial state effects are behind the suppression  is a cornerstone for the evidence that RHIC produced a ``new state of matter'' \cite{whitepaper1,whitepaper2,whitepaper3,whitepaper4}. 

 However, this consensus is undermined by the ambiguity over the correct approach to describing parton energy loss in terms of QCD.
Several models, treating jet-medium interactions through mutually inconsistent approximations, have been shown to successfully describe available experimental data \cite{amy2,glv3}.  

As a result, translating the jet energy loss observation into a quantitative  determination of the properties of the medium created at RHIC, such as an estimate of the initial partonic density or temperature, has been problematic.

Part of the reason for this ambiguity is that traditional jet identification and analysis tools are inapplicable to the heavy ion enviroenment, since
the background of hundreds of low energy tracks makes it impossible to distinguish every ``jet'' track from background low-momentum tracks independent of the jet.  

This makes  quantitative studies of jet broadening and jet structure in A-A collisions, capable of testing energy loss models, much more difficult, from a phenomenological and experimental point of view, than jet studies in e-e or p-p collisions.

QCD-based jet energy loss descriptions can be roughly divided in two classes:

i) {\em Thin plasmas}, where it is assumed that the system size is of the order of the parton's mean free path, and hence opacity (the mean number of hard collisions $\overline{n}$) becomes a suitable expansion parameter.   The dynamics within this approach can be extrapolated from an S-matrix like formalism.
The energy loss can therefore be calculated from a finite number of Feynman diagrams describing hard partonic collisions.  The most highly cited application of this approach is known as GLV (named after the authors)
\cite{loss3,glv1,glv2,glv3}.

ii) {\em Continuous absorption}, where it is assumed that the mean free path is much smaller than the path the jet has to traverse before hadronizing.
Hence, energy loss rate can be represented as occurring over ``many'' collisions occurring over each element of the parton's path within the medium.  The total energy loss can be 
calculated by integrating the resulting stochastic rate equation \cite{bdmps1,loss5}.  This approach was independently developed in \cite{bdmps1} (also named after the authors, BDMPS) and \cite{loss5}.
Recently, it was extended to make direct contact with finite temperature QCD via the so-called ``AMY'' formalism \cite{amy}

Note that the opacity parameter within the GLV formalism  is related to the partonic density in BDMPS and the medium temperature of AMY, since the mean free path parameter $\lambda_g$ can be related to the length of the parton's in-medium path $L$, the interaction
cross-section $\sigma$ and the partonic density $n$ via
\begin{equation}
\lambda_g \sim L \overline{n} \sim \frac{1}{\sigma n}
\end{equation}
The density in an ideal gas of light particles is (for $n_F$ light flavors and three colors) is then given by \cite{jansbook}
\begin{equation}
n \sim \frac{4 \pi}{(2 \pi)^3} 2 \left( 6 + \frac{3}{4} n_F  \right)  T^3
\end{equation}
Similarly, the color screening mass, appearing as a cut-off in GLV minimum gluon virtuality \cite{glv1}, is related
to the medium temperature and the strong fine structure constant $\alpha_s$ via
\begin{equation}
\label{mud}
\mu^2=4 \pi \alpha_s T^2
\end{equation}
The crucial difference between the two approaches, used to devise our experimental observable, is that in the thin plasma approximation the probability of {\em no interactions} remains finite, even if higher order terms in opacity are considered.  The continuous absorption approach, on the other hand,  is based on a stochastic equation where the transition rates include infinite re-summations of diagrams. The probability of no parton-medium interaction within this ansatz is manifestly zero {\em by construction}.

It is apparent that the  probability density function of the hard parton's momentum as a function of distance traveled behaves in a very different way in the two approaches.  

 Within the GLV ansatz, it maintains a primary peak, centered around the same $p_T$ and having the same width as immediately after the initial scattering.   
The \textit{amplitude} of this primary peak decreases with time, and eventually goes to zero
when the length traversed is much larger than the mean free path (a regime where the ``thin plasma'' approximation breaks down), but it's \textit{position} and \textit{ width} should
{\em not} change beyond effects due to fragmentation and primary jets reinteraction (common to all systems).   As the interaction probability grows, a secondary peak develops.  
Since the collisions within the GLV formalism are typically hard, this peak should be well removed, in phase space, from the primary peak.

\begin{figure*}[t]
\hspace*{-0.5cm}
\psfig{width=8cm,figure=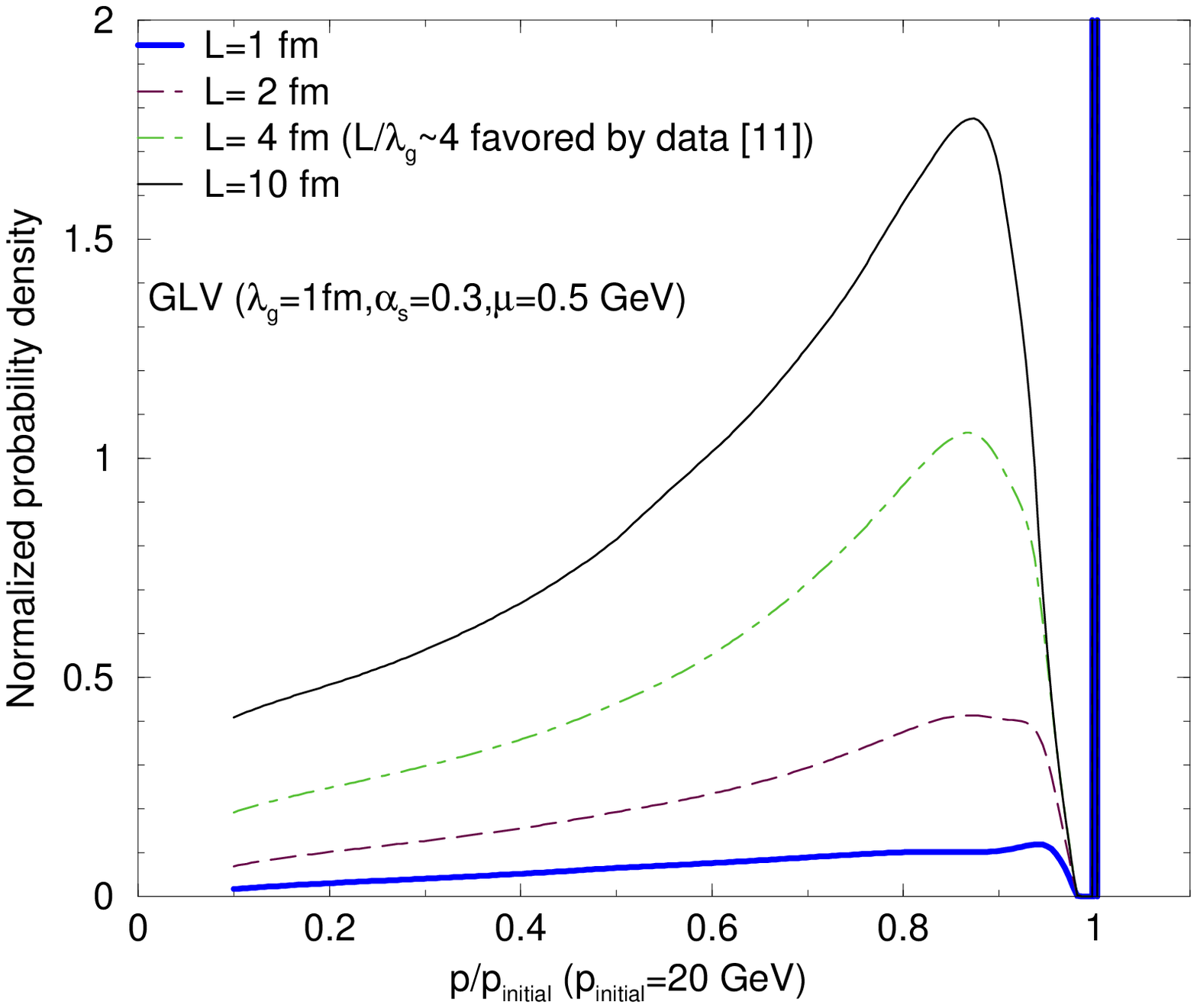}
\hspace*{2cm}
\psfig{width=8cm,figure=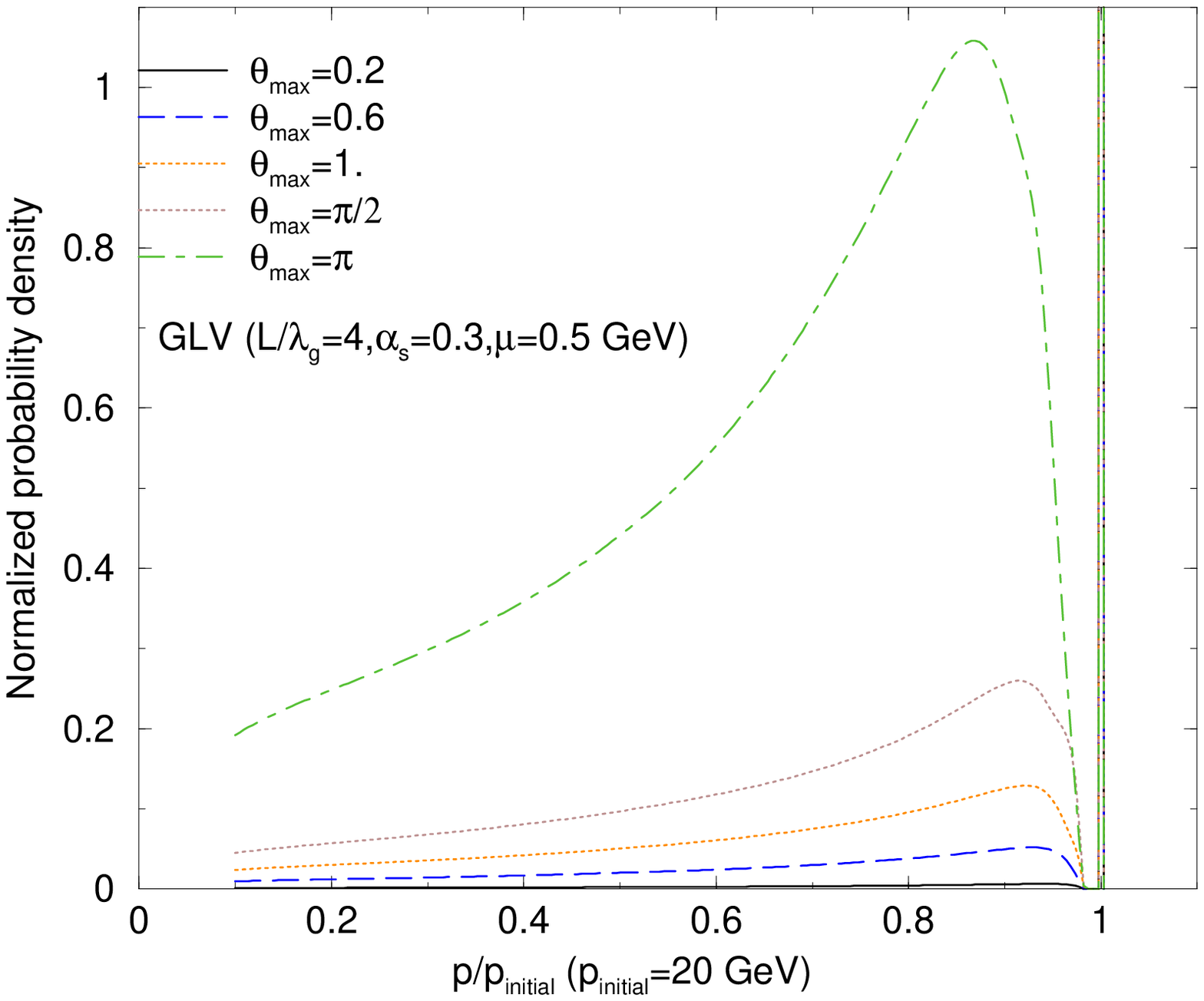}
\caption{\label{glvplot} 
Left panel:  probability density function for the partonic $p/p_{initial}$ calculated in the GLV formalism, to first order in opacity, as a function of distance traveled by the parton .
Right panel:  The secondary (rescattered) peak has been restricted to be within an angle $<\theta_{max}$ of the original trajectory.  See text around 
Eq. \ref{fig1right} for more discussion.
}
\end{figure*}

This picture is quantitatively investigated in Fig. \ref{glvplot}.  
In the left panel, we plot the probability density function for the surviving fraction of the initial momentum, calculated following the prescriptions in
\cite{glv2}
\begin{equation}
P_{GLV} \left( p \right) = \frac{1}{1+N} \left(\delta \left( p-  p_{ini} \right) + \frac{dI}{dp} \right)
\label{pglv}
\end{equation}
where  $p_{ini}$ is the initial parton momentum and $\frac{d I}{dp}$ is given, to first order in opacity, by
\begin{eqnarray}
\frac{dI}{d p}&=&\frac{2 C_{\rm R} \alpha_{\rm s}}{\pi}
\frac{p_{ini} L}{\lambda_{\rm g}} \, \int_0^1 \delta \left( x- 1+\frac{p}{p_{ini}} \right) {\rm d}x
\int_{\mu^2}^{k_{max}^2} \frac{{\rm d} {\bf k}^2_\perp}{{\bf k}^2_\perp} \nonumber \\
 & & \int_0^{\rm q_{\max}^2}
\frac{ {\rm d}^2{\bf q}_{\perp} \, \mu^2 }{\pi
({\bf q}_{\perp}^2 + \mu^2)^2 } \cdot
\frac{ 2\,{\bf k}_\perp \cdot {\bf q}_{\perp}
  ({\bf k} - {\bf q})_\perp^2  L^2}
{16x^2p_{ini}^2 \ + \ ({\bf k} - {\bf q})_\perp^4  L^2 }  
\label{inteq}
\end{eqnarray}
Here the $k_{max}^2$ is given by kinematic constraints
\[\  k_{max}^2={\rm min} \left[ 4 p_{ini}^2 x^2,4 p_{ini}^2 x(1-x) \right] \]
and N is a normalization constant to ensure the probability density function is correctly normalized
\begin{equation}
N = \int_\mu^{p_{mini}} \frac{dI}{d p} dp
\end{equation}
  As can be seen, re-scattering manifests itself by a secondary peak, whose area saturates at unity (probability of collision$=1$) as the size of the system becomes significantly larger than the mean free path. Multiple interactions (beyond first order in opacity) will change the secondary peak (Eq. \ref{inteq}) but, 
due to the ``thin plasma'' approximation, can not affect the center and the width of the primary peak (the $\delta$-function in Eq. \ref{pglv}).

The right panel of Fig \ref{glvplot} shows the same distribution where the integral in Eq. \ref{inteq} is restricted to a small angle between the medium parton ${\bf k}$ and the virtual gluon ${\bf q}$
\begin{equation}
\label{fig1right}
  |\theta_{k}-\theta_q| < \theta_{max} 
\end{equation}
and only the full angular distribution is normalized to unity.
It shows that the secondary peak is mostly at a large angle w.r.t. the original jet, significantly larger than a sensible experimentally defined jet cone.

Hence, if only hard particles within a jet-cone angle of $\sim 0.2$ w.r.t. the original parton direction are measured, it should be the case that either
{\em no} jet particles are found (the jet has been completely absorbed by the medium) or a jet that has approximately the original parton's momentum.

Within heavy ion collisions, of course, fragmentation, together with background tracks, makes
the definition of ``jet particle'' experimentally ambiguous. Furthermore, if we only observe one jet, we do not know the initial momentum of the original parton. 

  If, however, we observe a pair of azimuthally correlated jets, the original parton momentum can be inferred by the higher momentum ``near-side''
jet (probably close to the surface, due to the high opacity of the medium).    Provided the parton hadronizes outside the medium,
the broadening and hadronic momentum distribution due to fragmentation should be similar to that in d-Au collisions. 

Hence, the GLV approach predicts that, while the {\em probability} of having two azimuthally correlated jets {\em decreases}, {\em once they are found} they {\em also} maintain the momentum correlation expected from d-Au collisions.

The picture in the BDMPS/AMY formalism if very different.  The stochastic evolution ensures that, instead of a secondary peak developing, the primary peak broadens and eventually disappears  as the jet parton loses it's high momentum and becomes part of the medium.  
The fluctuation in the jet's momentum, therefore, is expected to increase with the distance traversed in the medium by the parton.  

\begin{figure*}[t]
\hspace*{2cm}
\psfig{width=8cm,figure=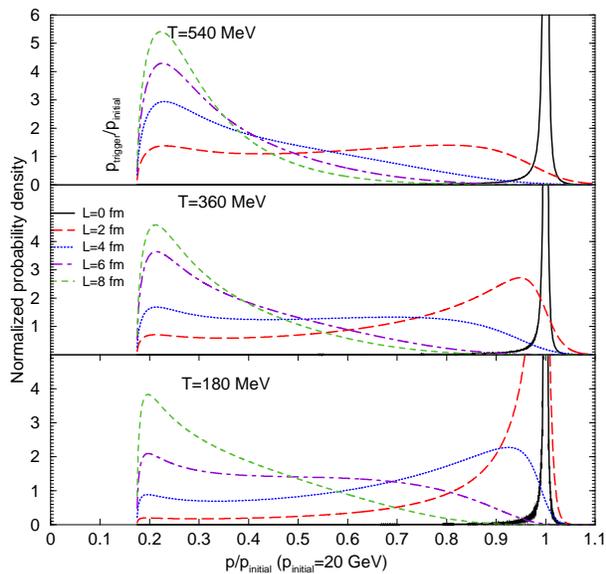}
\caption{\label{amyplot} 
probability density function of $p/p_{initial}$ as a function of distance traveled by the parton with the AMY formalism, for a variety of temperatures (particle densities).  All distributions are normalized to one, and require that $p_T>p_{Tmin}$ to avoid in-medium contamination.  The unphysical peak that develops at late
times at $p_{Tmin}$ is a consequence of this requirement.
}
\end{figure*}

Fig. \ref{amyplot} shows the fraction of the original momentum plotted against traversed length for a variety of medium temperatures and system sizes.
The calculation was done using a rate equation \cite{amy2,moorepriv}, where the
approximate solution is given by
\begin{eqnarray}
P_{AMY}(p, t) \approx \phantom{AAAAAAAAAAAAAAAAAAAAAAAA} \nonumber \\
\int d\epsilon\, D^{q} (\epsilon, p, t)\, P_{AMY}^q (p + \epsilon,0)+ D^{g} (\epsilon, p, t)\, P_{AMY}^g (p+\epsilon,0)  \, 
\label{eq:simple_sol}
\end{eqnarray}
where the Kernel is related to the transition rates\\ $\Gamma(p,\omega,t)$ calculated from finite-temperature QCD within the AMY formalism \cite{amy}
\begin{eqnarray}
D^{q/g}(\epsilon, p, t)= \exp\left[-\int_{-\infty}^{\infty} d\omega\, \right] \times \int_0^t dt'\, \Gamma(p, \omega, t')  \nonumber \\
\times \sum_{n=0}^\infty
{1\over n!}
\left[
\prod_{i=1}^n\int_{-\infty}^\infty d\omega_i\,
\int_0^t dt'\, \Gamma(p, \omega, t') 
\delta\left(\epsilon - \sum_{i=1}^n\omega_i\right)
\right]
\label{eq:Depsilon}
\end{eqnarray}
 the initial condition is given by a well-defined initial quark momentum
\begin{eqnarray}
P_{AMY}^q \left(p, t=0 \right) &=& \delta \left(p-p_{ini} \right)
\end{eqnarray}
 In addition, to prevent parton-medium ambiguity, we required
that the parton has an energy above a minimum initial $p_T$ of 2 GeV.

It can be seen that the initially sharp probability peak broadens and reaches a maximum width.   

Subsequently, the probability density function concentrates around the 
minimum $p_T$:  This peak is a consequence of our introduction of a minimum $p_T$ in the probability density function. It is not expected to be physical but is rather
an artifact of our ``jet'' definition.  It's appearance coincides with the parton becoming part of the medium, and the jet becoming unobservable.

As gluon splitting at small angles dominates in the AMY approach \cite{amy,amy2,moorepriv}, the broadening seen in Fig. \ref{amyplot} is independent of the jet-cone angle chosen, provided this angle is large enough to capture the whole jet and small enough to keep uncorrelated jets out.
therefore, a jet-cone angle similar to that defined within the GLV approach will also work here.

Thus, we have shown a qualitative difference between GLV and AMY energy loss Ansatzes.  Within GLV, the jet is absorbed through a decrease of it's {\em amplitude}, as expected from an approach derived via an ``S-Matrix'' like framework.  Within BDMPS/AMY, the jet disappears after an initial {\em broadening}, again as expected from the stochastic nature of the underlying ansatz.

These considerations lead us to propose the following observable:   We trigger on 
\begin{itemize}
\item An near-side hard particle with momentum $P_{Tnear}$
\item A away-side hard particle with a smaller but still jet-like momentum $P_{trigger}<P_{Taway}<P_{Tnear}$, where 
\[\  \pi-\delta \theta/2 < \theta_{near}-\theta_{away} < \pi+\delta \theta/2  \]
\end{itemize}
we compute total hard momentum in a cone with angle $\delta \theta$, defined as
\begin{equation}
\label{defnear}
P_{tot}^{near}(p_{Tmin},\delta \theta) = \sum_j^{p_{T}^j>P_{trigger},\left| \theta_j-\theta_{near} \right| <\delta \theta} P_{T j}
\end{equation}
we then calculate the equivalent away-side.
\begin{equation}
\label{defaway}
P_{tot}^{away}(p_{Tmin},\delta \theta) = \sum_j^{p_{T}^j>P_{trigger},\left| \theta_j-\theta_{away} \right| <\delta \theta} P_{T j}
\end{equation}
(note that the number of such particles $j$ in each case may well be 1)

Our observables are then the average and event-by-event fluctuation of z defined as 
\begin{equation}
\label{defz}
 z=P_{tot}^{away}/P_{tot}^{near}
\end{equation}
and it's 
 dependence on system size (centrality and nucleus type). Since we are using $P_{tot}^{near}$ as a gauge for the initial momentum of the parton, we require that $z<1$.
Hence, if in an event $z>1$  (an unlikely but possible outcome) this variable should be flipped ($z \rightarrow z^{-1}$) before being included in the statistics.

The considerations of the previous paragraphs show that the ``thin plasma'' approximation demands that
\begin{equation}
\ave{ z}_{A-A} \sim \ave{ z}_{d-Au}
\label{glvpheno1}
\end{equation}
\begin{equation}
\ave{ \left(\Delta z\right)^2}_{A-A} \sim \ave{ \left(\Delta z\right)^2}_{d-Au}
\label{glvpheno2}
\end{equation}
(where  $\ave{(\Delta X)^2} = \ave{X^2} - \ave{X}^2$)

Furthermore, GLV predicts no dependence with collision system size of either the average or the fluctuation in $z$.
A higher parton density/opacity will make collecting the statistical sample required for measuring these observables harder, but {\em once such a sample is obtained}
the results should follow the scaling of Eqs. \ref{glvpheno1} and \ref{glvpheno2} at any finite opacity. 

If, on the other hand, 
\begin{eqnarray}
\ave{ z}_{A-A} &<&\ave{ z}_{d-Au}  \\
\ave{ \left(\Delta z\right)^2}_{A-A}& >& \ave{ \left(\Delta z\right)^2}_{d-Au}
\end{eqnarray}
and $\ave{ \left(\Delta z\right)^2}_{A-A}$ exhibits a strong dependence on system size
(with a maximum in system size similar to the one present in Fig. \ref{amyplot})
than an AMY/BDMPS like ansatz is more appropriate for describing jet energy loss within the system.  
In this case, the system size where the maximum $\ave{ \left(\Delta z\right)^2}_{A-A}$ is observed can
be related to the temperature/partonic density of the created medium.

The leading source of systematic error for this measurement is the presence of {\em uncorrelated} jets within the system.
Hence, for this method to work, the important criterion in the definition of $P_{trigger}$ is to make sure that the probability of having {\em uncorrelated} 
hard particles in the same events with $P>P_{trigger}$ is small, so
\begin{equation}
\ave{ \left(\Delta z\right)^2} \simeq \ave{ \left( z\right)^2}
\end{equation}

To investigate the suitable $p_{trigger}$ and $\delta \theta$ for this measurement, we have analyzed the observables of interest using HIJING \cite{hijing}.
\begin{figure*}[t]
\centering
\hspace*{-0.5cm}
\psfig{width=8cm,figure=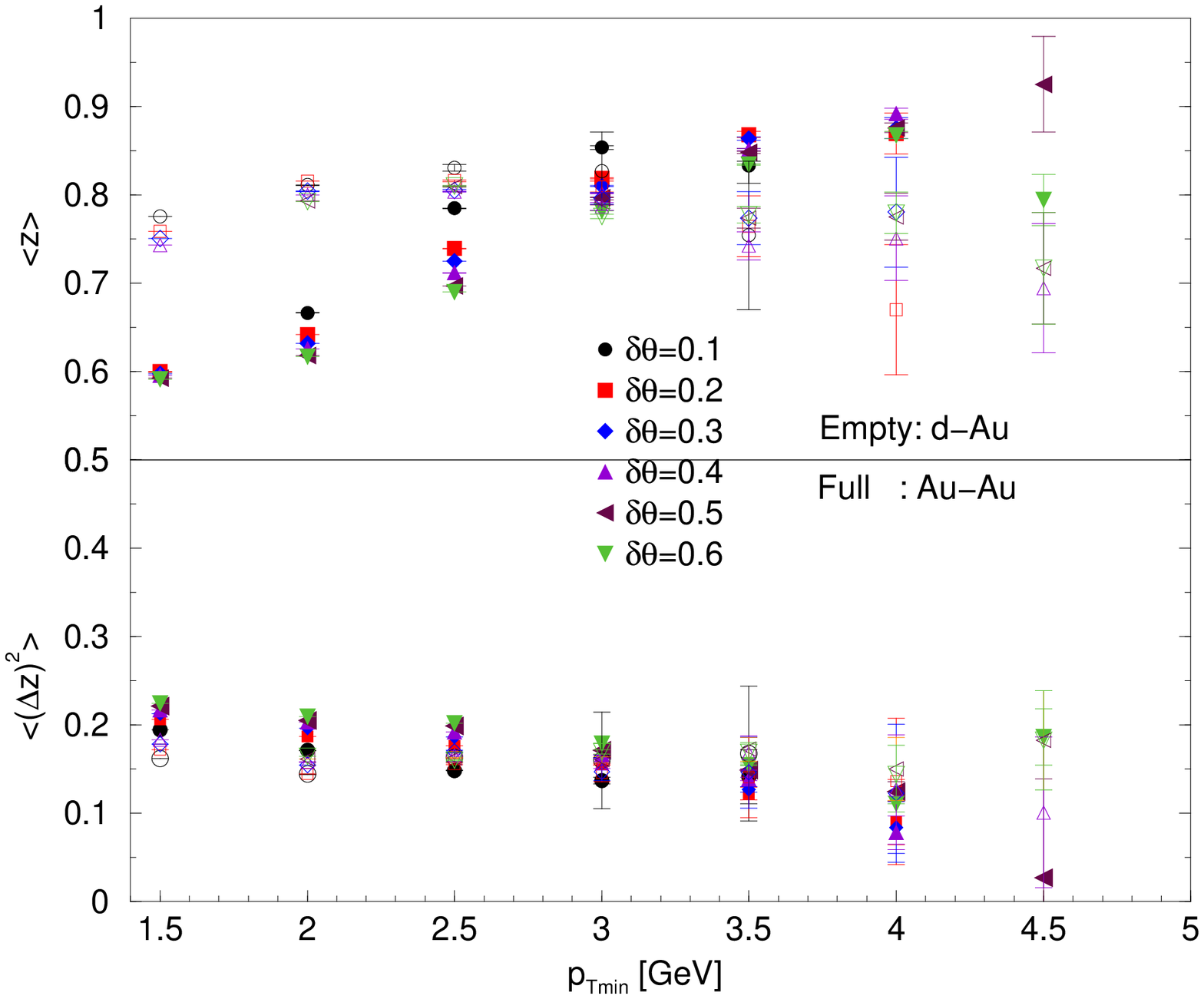}
\hspace*{2cm}
\psfig{width=8cm,figure=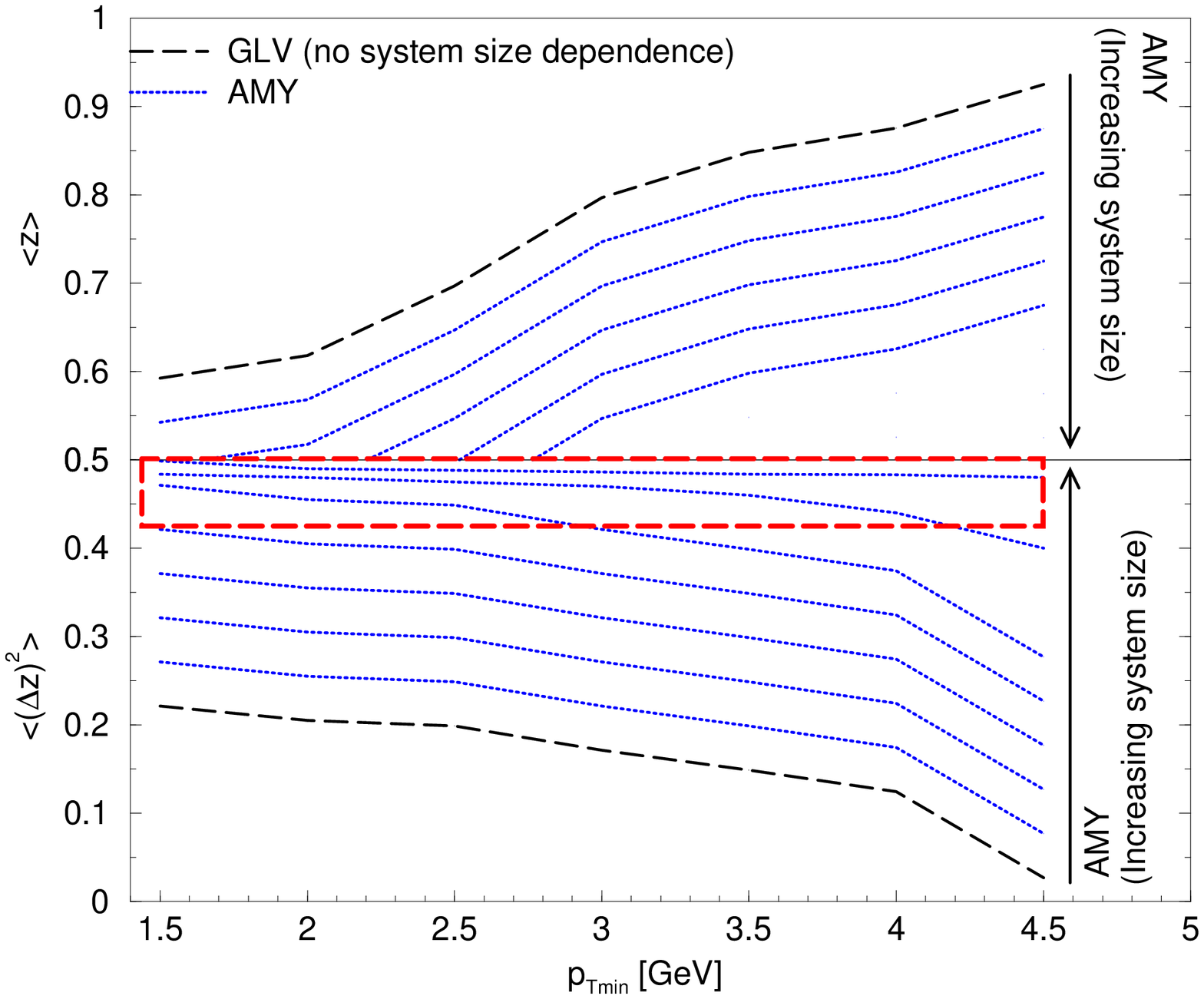}
\caption{\label{pp0_plot} Left panel: $\ave{ z}$ and $\ave{ \left(\Delta z\right)^2}$ calculated with HIJING and plotted for a variety of jet cone angles $\delta \theta$ and trigger momenta $p_{Tmin}$ (See eq. \ref{defz}, \ref{defnear} and \ref{defaway} for definition).  Empty symbols refer to d-Au, while full symbols are Au-Au.  Errors were calculated using the Jackknife method \cite{jackknife}.
Right panel: A schematic representation of the expectation for these observables in the GLV and AMY Ansatzes.  The dashed-line box in the bottom panel shows the system size of $\ave{ \left(\Delta z\right)^2}$ saturation, related to the medium's partonic density }
\end{figure*}

The result is shown in Fig. \ref{pp0_plot} for a variety of jet cone angles and trigger momenta.   As can be seen, if the jet cone angle is $< 0.6$ radians and $p_{Tmin}<4.5$ GeV, the $\ave{ z}_{A-A}$ and $\ave{ \left(\Delta z\right)^2}_{A-A}$ converge independently
of the precise jet cone angle value.  

The appearance of a strong jet cone angle dependence of $\ave{z},\ave{ \left(\Delta z\right)^2}_{A-A}$ at $p_{Tmin}=4.5$ GeV helps define an appropriate upper limit for
$p_{Tmin}$ (as per Eqs. \ref{defnear} and \ref{defaway}).  At that $p_{Tmin}$, the probability density for $z$ is dominated by the 
contamination of the chosen jet by independent hard processes (neighboring jets).   Of course, HIJING does not include the
significant quenching observed at RHIC, so the $p_{Tmin}$ limit can actually be pushed forward when analyzing experimental data.  

The onset of BDMPS/AMY regime should manifest itself by the broadening of the $z$ distribution.
Such broadening should increase $\ave{ \left(\Delta z\right)^2}_{A-A}$ for at least a sub-set of system size bins, up to a maximum value, fixed by the definition of $z$ to $\sim 0.5$.   

As Fig. \ref{amyplot} shows, the system size at which $\ave{ \left(\Delta z\right)^2}_{A-A}$ saturates to $\sim 0.5$  can be directly related to partonic
density (or temperature in the AMY approach). For T=540 MeV it's $\sim 2$ fm, for T=360 MeV it is $\sim$ 4 fm and for $T=180$ MeV it is $\sim 6$ fm.     
Hence, measuring this system size experimentally would remove the density/size ambiguity inherent in the interpretation of {\em average} jet suppression measurements.

Such a measurement should be experimentally feasible:  At the desired system size (be it centrality or colliding nucleus type), angular correlation of the produced jets should be nearly as high as in p-p and d-Au systems,
Since the distance traversed by the parton is not enough for a full in-medium absorption (no low-$p_T$ peak is in evidence for the critical system size in Fig. \ref{amyplot}).
  
However, $\ave{ \left(\Delta z\right)^2}_{A-A}$ will be  significantly above  $\ave{ \left(\Delta z\right)^2}_{d-Au}$ expectation.

Fig. \ref{pp0_plot} (right panel) also shows $\ave{ z}_{d-Au}$ and $\ave{ \left(\Delta z\right)^2}_{d-Au}$.   

While a $\sim 30 \%$ shift is observed between $\ave{z}_{A-A}$ and $\ave{z}_{d-Au}$ (expected and understood through the Cronin effect \cite{cronin}),  $\ave{ \left(\Delta z\right)^2}_{d-Au} \approx \ave{ \left(\Delta z\right)^2}_{Au-Au}  $ for {\em all} $p_{Tmin}$ triggers under consideration.   This 
small system size dependence (noted previously \cite{leonidov}) underscores the potential of $\ave{ \left(\Delta z\right)^2}$ as an experimental probe for the mechanism of energy loss.

We have verified that
these results are independent of the inclusion of $\ave{k_T}$ broadening \cite{hijingkt} within HIJING, but depend significantly on the inclusion of mini-jets.  Turning off the mini-jets results in a considerable decrease of $\ave{ \left(\Delta z\right)^2}_{d-Au}$, resulting in a $\sim 50\%$ difference w.r.t. $ \ave{ \left(\Delta z\right)^2}_{Au-Au}  $, as well as an expected reduction of the correlated jet sample.   
The divergence between  $\ave{ \left(\Delta z\right)^2}_{d-Au}$ and $ \ave{ \left(\Delta z\right)^2}_{Au-Au}$ in this regime is not surprising, since within HIJING the dominant contributing factor to 
 $\ave{ \left(\Delta z\right)^2}$ is the generation of mini-jets by the propagating jet.
We have also calculated $\ave{z}_{p-p}$ and  $\ave{ \left(\Delta z\right)^2}_{p-p}$, and the result is predictably indistinguishable up to statistical error from d-Au.

In conclusion, we have illustrated the qualitative difference between GLV and BDMPS/AMY energy loss dynamics, the first being based on the {\em decrease} of the initial jet momentum peak, while the second predicting a {\em broadening} of the peak. 
We have argued that this difference should result in different scaling, with system size (centrality or nucleus type), of the mean and variance of the variable
$z=P_T^{away}/P_T^{near}$.   We have shown that in a hadronic enviroenment, as simulated by HIJING, a sensible combination of $p_T$ trigger and jet-cone angle will give a $P_T^{away}/P_T^{near}$
independent of jet-cone angle and momentum trigger systematics and close to the ``thin plasma'' (GLV) limit, thus providing a reliable way to search for the onset of BDMPS/AMY broadening.  We have also argued that the dependence of this broadening on system size can be related to the initial partonic density of the fireball.   We hope that experimental measurements of $\ave{z}$ and $\ave{(\Delta z)^2}$ will be forthcoming, and will help us learn more about the mechanism of hadronic energy loss.


We thank  Guy Moore for having supplied the computer program used to calculate results in the AMY formalism.
We also thank Guy Moore and Simon Turbide for helpful discussions and explanations.
S.J.~thanks RIKEN BNL Center
and U.S. Department of Energy [DE-AC02-98CH10886] for
providing facilities essential for the completion of this work.
Work supported in part by  grants from
the U.S. Department of Energy  (J.R. by DE-FG02-04ER41318),
the Natural Sciences and Engineering research
council of Canada, the Fonds Nature et Technologies of Quebec.
GT thanks the Tomlinson foundation for the support provided.

\vskip 0.3cm

\end{document}